%% file: article.tex
\documentclass{article}
\usepackage{amsmath,graphicx}
\usepackage[preprint]{spconf}
\usepackage{subcaption}
\usepackage{amssymb}
\usepackage{bm}

\usepackage{siunitx}
\usepackage{url}
\usepackage{tikz}
\usepackage{tikz, tikzscale}
\usetikzlibrary{matrix,chains,positioning,decorations.pathreplacing,arrows,fadings}

\usepackage{pgfplotstable}
\usepackage{pgfplots}
\usepgfplotslibrary{units}
\usetikzlibrary{pgfplots.statistics, pgfplots.colorbrewer}
\usepackage{siunitx}
\pgfplotsset{compat=1.9}
\makeatletter
\pgfplotsset{
    unit code/.code 2 args=
    \begingroup
    \protected@edef\x{\endgroup\si{#2}}\x
}

\usepackage{cite}
\def\x{{\mathbf x}}

\usepackage[acronym]{glossaries}
\makeglossaries
\newacronym{rir}{RIR}{Room Impulse Response}
\newacronym{rirs}{RIRs}{Room Impulse Responses}
\newacronym{toa}{ToA}{Time of Arrival}
\newacronym{sofa}{SOFA}{Spatially Oriented Format for Acoustics}

\title{HOMULA-RIR: A Room Impulse Response Dataset for Teleconferencing and Spatial Audio Applications Acquired Through Higher-Order Microphones and Uniform Linear Microphone Arrays}
\name{\begin{minipage}{\textwidth}\centering Federico Miotello, Paolo Ostan, Mirco Pezzoli, Luca Comanducci,\\Alberto Bernardini, Fabio Antonacci, Augusto Sarti\end{minipage}\thanks{This work has been funded by "REPERTORIUM project. Grant agreement number 101095065. Horizon Europe. Cluster II. Culture, Creativity and Inclusive Society. Call HORIZON-CL2-2022-HERITAGE-01-02."}\thanks{This work was supported by the Italian Ministry of University and Research (MUR) under the National Recovery and Resilience Plan (NRRP), and by the European Union (EU) under the NextGenerationEU project.}}
\address{Dipartimento di Elettronica, Informazione e Bioingegneria, Politecnico di Milano, Milan, Italy}
\toappear{\begin{minipage}{\textwidth}\footnotesize© 20XX IEEE. Personal use of this material is permitted. Permission from IEEE must be obtained for all other uses, in any current or future media, including reprinting/republishing this material for advertising or promotional purposes, creating new collective works, for resale or redistribution to servers or lists, or reuse of any copyrighted component of this work in other works.\end{minipage}}
\begin{document}
\ninept
\maketitle
\begin{abstract}
In this paper, we present HOMULA-RIR, a dataset of room impulse responses (RIRs) acquired using both higher-order microphones (HOMs) and a uniform linear array (ULA), in order to model a remote attendance teleconferencing scenario.
Specifically, measurements were performed in a seminar room, where a $64$-microphone ULA was used as a multichannel audio acquisition system in the proximity of the speakers, while HOMs were used to model $25$ attendees actually present in the seminar room.
The HOMs cover a wide area of the room, making the dataset suitable also for applications of virtual acoustics.
Through the measurement of the reverberation time and clarity index, and sample applications such as source localization and separation we demonstrate the effectiveness of the HOMULA-RIR dataset.
\end{abstract}
%
\input{sections/01_introduction}
\input{sections/02_description}
\input{sections/03_evaluation}
\input{sections/04_application}
\input{sections/05_conclusion}
\input{sections/06_ack}
\bibliographystyle{ieeetr}
{\footnotesize
\bibliography{refs}}

\end{document}

%% file: sections/01_introduction.tex
\section{Introduction}\label{sec:introduction}
In recent years, teleconferencing platforms have become part of daily lives of most people, especially after the COVID-19 pandemic.
Applications like source separation~\cite{olivieri2023real}, speech enhancement~\cite{hsu2022learning,eskimez2022personalized} or echo-canceling~\cite{sridhar2021icassp}, dereverberation~\cite{chen2023learning} or audio packet loss concealment \cite{diener2022interspeech, mezza2023hybrid, miotello2023deep} are customarily used in teleconferencing software.
Moreover, with the growing interest within augmented and virtual reality contexts, an increasing number of platforms, including those for teleconferencing, are incorporating spatial audio features.
To this end, tasks such as sound field reconstruction \cite{pezzoli2020parametric, mccormack2022parametric, karakonstantis2023generative} have gained particular relevance, due to their crucial role in enabling applications like navigable audio.
Nevertheless, to accurately assess the performance of these methods in real-world scenarios, it is necessary to test them using data measured in real environments.
Additionally, more and more approaches nowadays heavily rely on machine learning or other data-driven algorithms, that thus need large amounts of data for training and validation.
For these reasons, several \acrfull{rir} datasets are present in the literature.
In~\cite{hadad2014multichannel} multichannel responses were measured in a room with variable reverberation levels, with the aim of evaluating source separation techniques, while in~\cite{koyama2021meshrir} \acrshort{rirs} in low-reverberation time rooms were measured to test sound zone control and reconstruction methods. The Multi-arraY Room Acoustic Database (MYRiAD) dataset~\cite{dietzen2023myriad} was created by acquiring measurements using several microphone configurations such as in-ear omnidirectional microphones in a dummy head, circular arrays and behind-the-ear arrays in two recording scenarios.
\begin{figure}[t]
    \centering
    \frame{\includegraphics[width=\columnwidth]{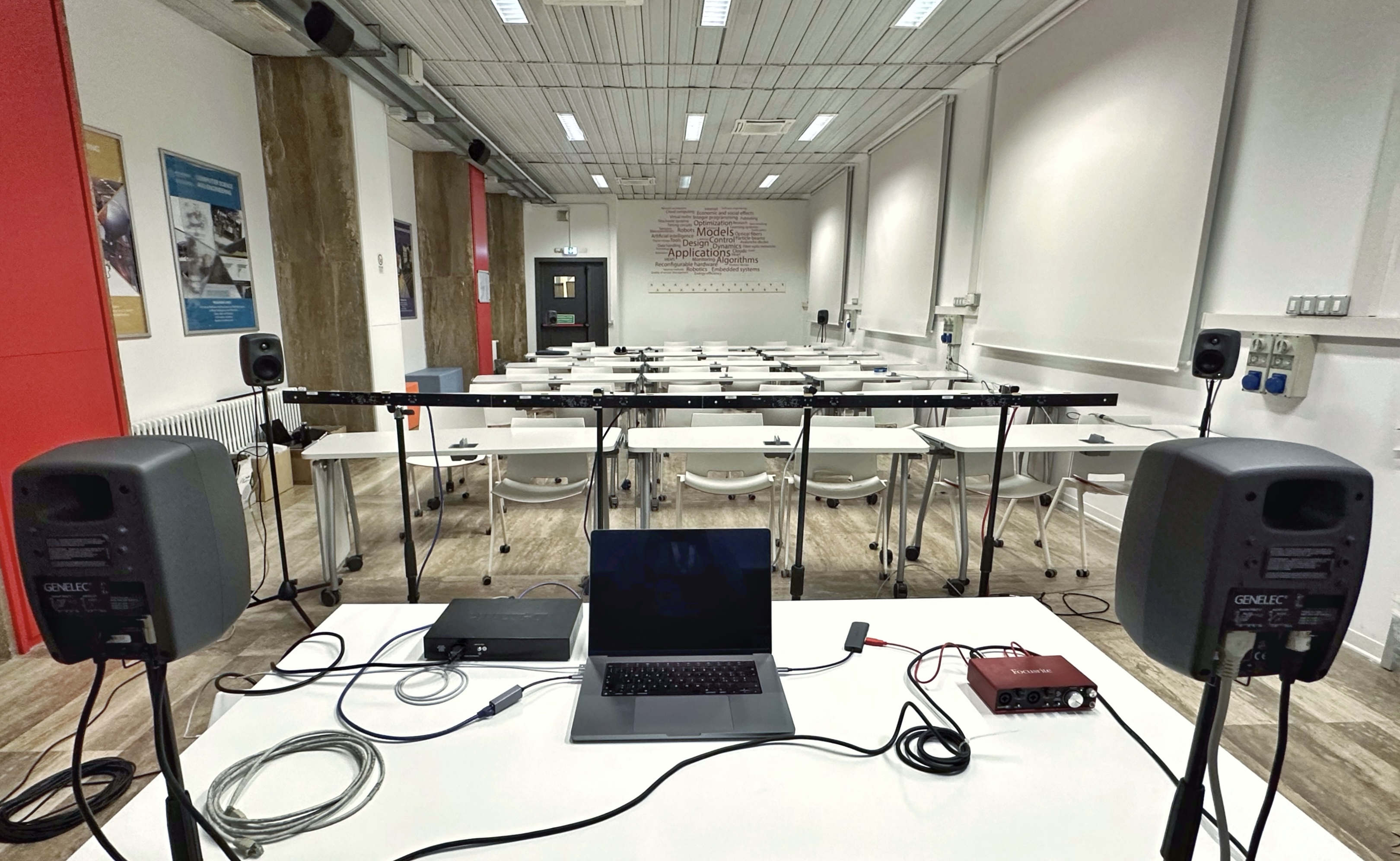}}
    \caption{Photograph of ``Schiavoni room'' taken during the measurement session.}
    \label{fig:room}
\end{figure}
Several \acrshort{rir} datasets acquired through higher-order microphones (HOMs) have also been released.
In~\cite{stewart2010database} measurements were acquired in three rooms, each with a static source position and more than one-hundred receivers, using both omnidirectional and B-format microphones. 
A dataset of six degrees-of-freedom \acrshort{rirs} in controlled and empty rooms, using different reverberation levels was presented in~\cite{choi2023six}, while the Motus dataset~\cite{gotz2021dataset} was created by acquiring Ambisonic \cite{zotter2019ambisonics} \acrshort{rirs} in a single room and varying the furniture position. 
\begin{figure*}[t]
    \centering
    \includegraphics[width=0.8\textwidth]{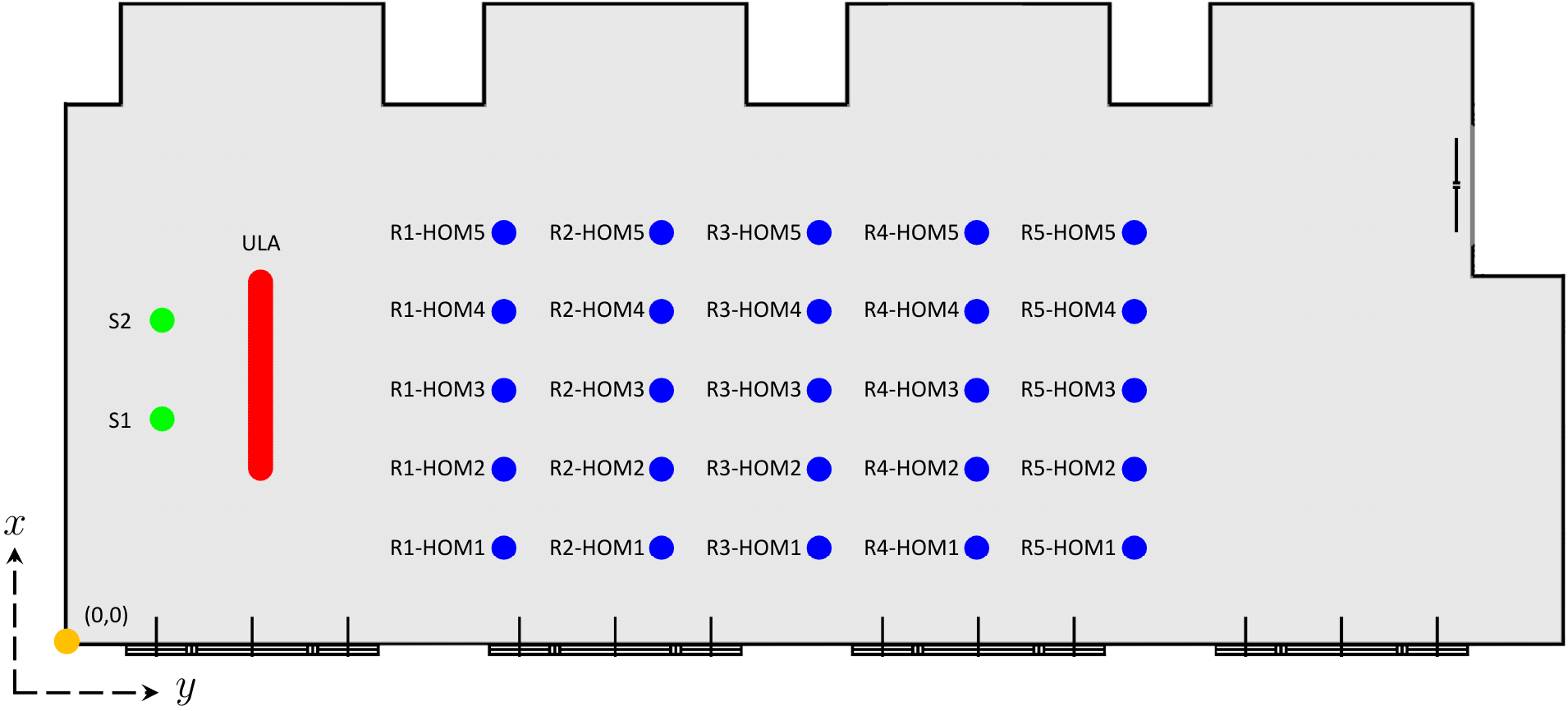}
    \caption{Floor plan of Schiavoni seminar room. Sources are depicted in green, ULAs in red, and HOMs in blue. Spatial measurements are referenced to the orange marker denoting the origin point.}
    \label{fig:room_plan}
\end{figure*}

In this paper, we present HOMULA-RIR, a complementary dataset of \acrshort{rirs} acquired in a real environment using a hybrid setup, with the objective of representing a realistic teleconferencing scenario. 
Specifically, we have deployed a linear microphone array to simulate the acquisition of the main speaker, e.g., a lecturer, by a teleconferencing audio system; and HOMs densely sampling the listeners position within the room. 
Measurements were performed in ``Schiavoni room'' located at Dipartimento di Elettronica, Informazione e Bioingegneria of Politecnico di Milano in Milan, Italy.
The seminar room is named after Prof. Emer. Nicola Schiavoni and it is employed for lectures and teleconferences by the staff of the Politecnico di Milano. 
After the acquisition of the \acrshort{rirs} and a geometric calibration of the arrays based on acoustic measures, we estimate the reverberation time and the clarity of the room.
Moreover, in order to validate the HOMULA-RIR dataset, we test it under two sample applications, namely source localization and source separation, demonstrating its effectiveness.
The rest of the paper is organized as follows.
In Sec.~\ref{sec:dataset_description} we describe the dataset in terms of environment and setup, while in Sec.~\ref{sec:evaluation} we describe some objective measurements related to the environment conditions. In Sec.~\ref{subsec:application} we present two sample applications.
Finally, in Sec.~\ref{sec:conclusion} we draw some conclusions.
The dataset is freely available at \url{https://doi.org/10.5281/zenodo.10479726}. 

%% file: sections/02_description.tex
\section{Dataset description}\label{sec:dataset_description}
The proposed dataset is composed of a collection of RIRs, measured in a seminar room covering a wide area thereof.
Considering the designated use of the dataset for teleconferencing and spatial audio applications, we chose to employ a hybrid setup by capturing \acrshort{rirs} using both uniform linear arrays (ULAs) and HOMs.
More specifically, we opted to position the ULAs in front of the desk, typically where one or more lecturers are located during a presentation, to emulate a teleconferencing system capturing the sound in the proximity of the source.
The HOMs, instead, are positioned in correspondence of the listeners' seats, to replicate the listening perspective of the attendees.
The \acrshort{rirs} have been recorded using logarithmic sine sweeps ranging from \SI{50}{\hertz} to \SI{22}{\kilo\hertz}, each lasting \SI{10}{\second} and sampled at \SI{48}{\kilo\hertz}.
The room responses have been acquired using Reaper as Digital Audio Workstation and all the audio streams were routed through the Dante\texttrademark \, Controller to a laptop operating a Dante\texttrademark \, Virtual Soundcard.

\subsection{Room conditions}
The room in which we performed the measurements, shown in Figure~\ref{fig:room}, is typically used for frontal lessons and seminars and thus is furnished with tables and chairs for both lecturers and the audience.
To preserve and capture the acoustic properties of the actually used space, we opted to keep the furniture during the collection of the \acrshort{rirs}.
The room is \SI{14.52}{\meter} long, \SI{5.46}{\meter} wide and \SI{3.38}{\meter} high. 
The floormap is shown in Figure \ref{fig:room_plan}.
It features concrete and tile surfaces throughout, with windows covered by heavy curtains.

\subsection{Sources setup}
We considered two sources located behind the main desk in the room, specifically employing two Genelec 8020D loudspeakers\footnote{\url{https://www.genelec.com/8020d}}, which can be seen in the foreground in Figure~\ref{fig:room}.
The intention is to replicate the scenario of two speakers or lecturers addressing an audience in the room.
With respect to the origin indicated in Figure \ref{fig:room_plan}, the sources are located at positions $S_1 = [\SI{2.28}{\meter} \; \SI{0.96}{\meter} \; \SI{1.20}{\meter}]^T$ and $S_2 = [\SI{3.28}{\meter} \; \SI{0.96}{\meter} \; \SI{1.20}{\meter}]^T$, considering the acoustic center indicated in the operating manual of the loudspeakers as measurement point.

\subsection{EStick setup}
As far as the ULAs are concerned, we used four co-linear EStick V3, made by Eventide Inc. in collaboration with Politecnico di Milano \cite{pezzoli2018a}.
ESticks are modular ULAs, and each unit consists of 16 omnidirectional MEMS microphones with a spacing of \SI{3}{\centi\meter} between them.
One advantage of the system is its versatility in rapidly deploying various linear and planar array configurations, accommodating up to 64 microphone elements.
For our particular setup, visible in Figure~\ref{fig:room}, we opted for a linear geometry to achieve a 64-microphone-long array, positioned in front of the main desk, at the same height $z = \SI{1.20}{\meter}$ of the sources.
The microphone signals are accessible through the integrated Dante\texttrademark \, connectivity, using a single CAT6 cable for each array, which provides both power and synchronization.
Even if the position of each capsule is known a priori, we have implemented a self-calibration procedure to localize the ULA within the room, as described in Section~\ref{sec:calibration}.

\subsection{Spatial Mic setup}
For the acquisition of the \acrshort{rirs} through HOMs, we adopted the Spatial Mic Dante by Voyage Audio\footnote{\url{https://voyage.audio/spatialmic}}.
The Spatial Mic is based on the geometry presented in \cite{benjamin2012second}, and it features 8 prepolarized condenser capsules, allowing a higher-order Ambisonics \cite{zotter2019ambisonics} encoding up to the $2^{\text{nd}}$ order.
Similarly to the ESticks, the Spatial Mic features integrated Dante\texttrademark \, connectivity, using a single CAT6 cable for each unit that provides both power and synchronization.
In the considered setup, we positioned the HOMs in correspondence of the audience seats, at the same height $z = \SI{1.20}{\meter}$ of the sources, covering $25$ different locations divided in $5$ rows.
With five devices at our disposal, we conducted the measurements five times, relocating the microphones while maintaining precise consistency in the surrounding environment and position of the sources.
The position of each capsule has been estimated through acoustic measures computed as described in Section \ref{sec:calibration}.

\subsection{Data format}
The released \acrshort{rirs} have been recorded at a sample rate of $\operatorname{fs} = \SI{48}{\kilo\hertz}$ and truncated to a duration of \SI{1}{\second}. They are provided as multichannel \texttt{wav} files, saved at \SI{32}{\bit} per sample.
\acrshort{rirs} of individual arrays are saved as separate files, following the naming convention: \texttt{rir-\textbf{source}-\textbf{array}.wav}.
Here, \texttt{\textbf{source}} can be either \texttt{S1} or \texttt{S2}, depending on the considered source, and \texttt{\textbf{array}} is an acronym representing a specific microphone array, as depicted in Figure \ref{fig:room_plan}.
The term \texttt{\textbf{array}} can take on either \texttt{ULA} for the ESticks measures, or a pair \texttt{\textbf{row}-\textbf{HOM}} for the Spatial Mics measures.
Specifically, $\text{\texttt{\textbf{row}}} = \{ \text{\texttt{R1, R2, R3, R4, R5}} \}$ designates the row where a particular Spatial Mic is positioned, and $\text{\texttt{\textbf{HOM}}} = \{ \text{\texttt{HOM1, HOM2, HOM3, HOM4, HOM5}}\}$ denotes a specific array within each row.
The positions of each capsule in every array are released as \texttt{csv} files, adopting the naming convention \texttt{pos-\textbf{array}.csv}, where \texttt{\textbf{array}} is the same acronym denoting a specific microphone array.
Additionally, the positions of the two sources are reported in the file \texttt{pos-sources.csv}.

%% file: sections/03_evaluation.tex
\section{Evaluation of measurements}\label{sec:evaluation}
\subsection{Calibration}\label{sec:calibration}
A calibration process is carried out to determine the absolute placement within the room of each capsule of the microphone arrays.
This allows us to provide acoustically precise positions, which are particularly relevant for many potential applications of the data set.
To infer each microphone's location, the calibration process uses four additional loudspeakers (Genelec 8020D), whose positions are known and which can be seen in the background in Figure \ref{fig:room}.

First, we measured the \acrshort{rirs} between each capsule of both the ESticks and the Spatial Mics, and the additional sources placed around the area in which the microphones are located.
The acquisition of the \acrshort{rirs} follows the same procedure detailed in Section \ref{sec:dataset_description}.
From each \acrshort{rir}, by identifying the first peak associated to the direct path, we compute the \acrfull{toa}, which is the time that it takes for a pressure wave to travel from a source to one of the microphone capsules.
Then, to estimate the unknown positions of each capsule $\mathbf{r}_m$, we solve an optimization least-squares problem in 2D as 
\begin{equation}
    (\mathbf{r}_m, d) = \mathop{\arg \min}\limits_{(\mathbf{r}_m, d)} \sum_{m, l} [
        \| \mathbf{r}_m - \mathbf{r}_l \| - (\tau_{m,l} \cdot c - d)
    ]^2,
\end{equation}
where $c$ is the speed of sound, $\mathbf{r}_l$ is the position of loudspeaker $l$, $\tau_{m,l}$ is the \acrshort{toa} between $\mathbf{r}_m$ and $\mathbf{r}_l$ and $d$ is an estimate of the delay caused by the acquisition system latency, expressed in meters.
\subsection{Reverberation time}
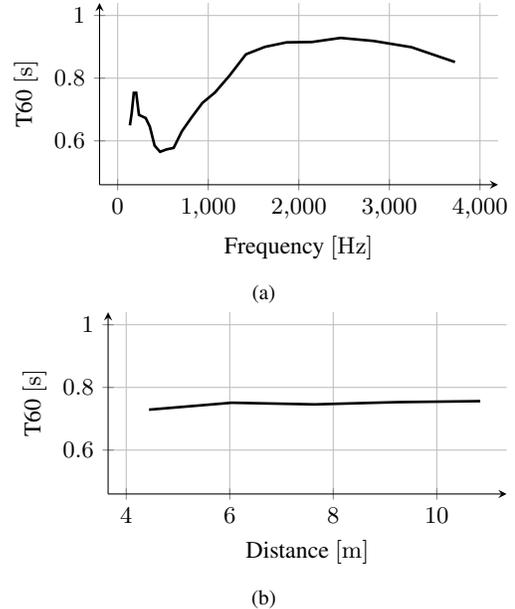
\begin{figure}[t]
     \centering
     \begin{subfigure}{\columnwidth}
        \centering
        \input{results/t60_freq}
        \caption{}
        \label{fig:t60_freq}
     \end{subfigure}
     \begin{subfigure}{\columnwidth}
         \centering
        \input{results/t60_dist}
        \caption{}
        \label{fig:t60_dist}
     \end{subfigure}
    \caption{Reverberation time as a function of (a) frequency and as a function of (b) distance from the sources.}
    \label{fig:t60}
\end{figure}
In order to characterize the acoustic properties of ``Schiavoni room'', we estimate the reverberation time $\operatorname{T60}$ of the environment.
In particular, we compute the average room $\operatorname{T60}$ as the mean $\operatorname{T60}$ measured from each \acrshort{rir} using Schroeder method, in the implementation provided by the \texttt{pyroomacoustics} package \cite{scheibler2018pyroomacoustics}.
In particular, we considered a \SI{30}{\decibel} decay in the energy decay curve to actually estimate the $\operatorname{T60}$.
Figure~\ref{fig:t60_freq} shows the variation of $\operatorname{T60}$ across third-octave frequency bands, spanning from \SI{125}{\hertz} to \SI{4000}{\hertz}.
Figure \ref{fig:t60_dist}, instead, depicts the relationship between $\operatorname{T60}$ and the absolute $y$ position in the room, providing insights into reverberation time dependence on the distance from the sources.
As expected, the reverberation time exhibits variability among different frequency bands, ranging from a minimum of \SI{0.56}{\second} to a maximum of \SI{0.93}{\second}, with a mean value of \SI{0.74}{\second}.
In contrast, there is no discernible dependence on distance, indicating that the reverberation time remains consistent throughout different locations within the room.
\subsection{Clarity index}
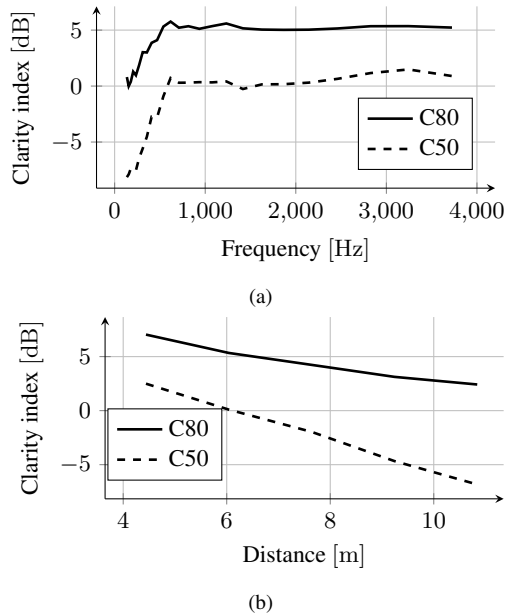
\begin{figure}[t]
     \centering
     \begin{subfigure}{\columnwidth}
        \centering
        \input{results/c80_freq}
        \caption{}
        \label{fig:c80_freq}
     \end{subfigure}
     \begin{subfigure}{\columnwidth}
         \centering
        \input{results/c80_dist}
        \caption{}
        \label{fig:c80_dist}
     \end{subfigure}
    \caption{Clarity index as a function of (a) frequency and as a function of (b) distance from the sources.}
    \label{fig:c80}
\end{figure}
The clarity index, as defined in ISO-3382-1 standard, serves as an estimation of perceived clarity within a room and depends on the energy ratio between the early and late parts of the \acrshort{rir} \cite{gotz2023online}.
Specifically, we calculated both $\operatorname{C50}$ and $\operatorname{C80}$ clarity indices, using the implementation available in the \texttt{python-acoustics} package.
Figure~\ref{fig:c80_freq} displays the variation of both metrics across third-octave frequency bands, spanning from \SI{125}{\hertz} to \SI{4000}{\hertz}.
Figure~\ref{fig:c80_dist}, instead, presents the relationship between clarity index and the absolute $y$ position in the room, enabling the exploration of its dependence on the distance from the sources.
Except for the lower frequencies, clarity remains relatively consistent across different frequency bands, converging towards average values of $\operatorname{C50} = \SI{0.5}{\decibel}$ and $\operatorname{C80} = \SI{5.4}{\decibel}$.
Conversely, when considering the dependence on distance, $\operatorname{C50}$ exhibits a decline from \SI{2.5}{\decibel} at the front row of the room to \SI{-6.8}{\decibel} at the back of the room, while $\operatorname{C80}$ exhibits a decline from \SI{7.0}{\decibel} to \SI{2.4}{\decibel}.

%% file: results/t60_freq.tex
\begin{tikzpicture}%
\begin{axis}[%
    xlabel={Frequency},%
    x unit = \hertz,%
    ylabel={T60},%
    y unit = \second, %
    xtick={0, 1000, 2000, 3000, 4000},
    xmin=0, xmax=4000,
    ymax=1, ymin=0.5,
    axis x line=bottom,%
    axis y line=left, %
    grid, %
    height=4cm, %
    width=0.8\columnwidth, %
    enlarge x limits=0.05,%
    enlarge y limits=0.08,%
    style={font=\normalsize},%
]

\addplot[mark=, line width=1pt] table {results/t60_freq.txt};%

\end{axis}%
\end{tikzpicture}

%% file: results/t60_dist.tex
\begin{tikzpicture}%
\begin{axis}[%
    xlabel={Distance},%
    x unit = \meter,%
    ylabel={T60},%
    y unit = \second, %
    xtick={4, 6, 8, 10},
    xmin=4, xmax=11,
    ymax=1, ymin=0.5,
    axis x line=bottom,%
    axis y line=left, %
    grid, %
    height=4cm, %
    width=0.8\columnwidth, %
    enlarge x limits=0.05,%
    enlarge y limits=0.08,%
    style={font=\normalsize},%
]

\addplot[mark=, line width=1pt] table {results/t60_dist.txt};%

\end{axis}%
\end{tikzpicture}

%% file: results/c80_freq.tex
\begin{tikzpicture}%
\begin{axis}[%
    xlabel={Frequency},%
    x unit = \hertz,%
    ylabel={Clarity index},%
    y unit = \decibel, %
    xtick={0, 1000, 2000, 3000, 4000},
    xmin=0, xmax=4000,
    ymax=6, ymin=-8,
    axis x line=bottom,%
    axis y line=left, %
    grid, %
    height=4cm, %
    width=0.8\columnwidth, %
    enlarge x limits=0.05,%
    enlarge y limits=0.08,%
    style={font=\normalsize},%
    legend style={at={(0.8,0.1)},anchor=south,font=\normalsize},%
]

\addplot[line width=1pt] table {results/c80_freq.txt};%
\addlegendentry{C80}
\addplot[dashed, line width=1pt] table {results/c50_freq.txt};%
\addlegendentry{C50}

\end{axis}%
\end{tikzpicture}

%% file: results/c80_dist.tex
\begin{tikzpicture}%
\begin{axis}[%
    xlabel={Distance},%
    x unit = \meter,%
    ylabel={Clarity index},%
    y unit = \decibel, %
    xtick={4, 6, 8, 10},
    xmin=4, xmax=11,
    ymax=7.5, ymin=-7,
    axis x line=bottom,%
    axis y line=left, %
    grid, %
    height=4cm, %
    width=0.8\columnwidth, %
    enlarge x limits=0.05,%
    enlarge y limits=0.08,%
    style={font=\normalsize},%
    legend style={at={(0.15,0.1)},anchor=south,font=\normalsize},%
]

\addplot[line width=1pt] table {results/c80_dist.txt};%
\addlegendentry{C80}
\addplot[dashed, line width=1pt] table {results/c50_dist.txt};%
\addlegendentry{C50}

\end{axis}%
\end{tikzpicture}

%% file: sections/04_application.tex
\section{Sample application}\label{subsec:application}
In order to validate the HOMULA-RIR dataset, we conducted various tests targeting two classical applications: blind source separation and source localization.
Specifically, blind source separation was employed to show the use of ULA signals, while source localization was used to validate the HOMs signals.

\subsection{Blind source separation}
The task of blind audio source separation involves extracting multiple unknown audio signals (sources) by processing their combined mixture.
This process is referred to as \textit{blind} because the algorithm has access only to the mixed signals and lacks information about the individual source signals.
To achieve this, we exploit the Ray-Space-Based Multichannel Nonnegative Matrix Factorization algorithm (RS-MNMF), originally proposed in \cite{pezzoli2021ray}.
The algorithm leverages the Ray Space Transform \cite{bianchi2016ray}, to project the microphone signals acquired by ULAs (the ESticks) into the Ray Space domain.
In such domain, the position of sources is encoded directly into the magnitude of the ray-space-transformed signals.
This results in an effective use of the spatial information present in the mixture and encoded in the ray space data, allowing for a direct application of the conventional multichannel NMF algorithm \cite{pezzoli2021ray}.

Results are computed in terms of three classic blind source separation metrics, namely Source to Distortion Ratio (SDR), Source to Interferences Ratio (SIR) and Sources to Artifacts Ratio (SAR) \cite{vincent2006performance}.
The tests were conducted considering \SI{3}{\second} long speech signals, and the average value of these metrics over all the microphone signals is taken into account.
It can be noticed that the results remain consistent for both examined sources, $S_1$ and $S_2$, indicating that the separation can be successfully performed considering either sources and independently from the locations.
Additionally, the obtained values align with those presented in \cite{pezzoli2021ray} using a similar setup.
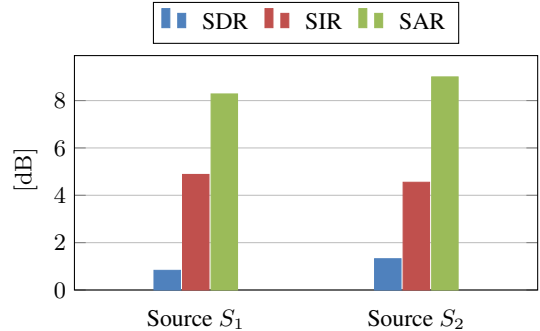
\begin{figure}[t]
    \centering
    \input{results/separation}
    \caption{Blind source separation results: SDR, SIR and SAR metrics for both considered sources.}
    \label{fig:separation}
\end{figure}
\subsection{Source localization}
In the context of acoustic signal processing, source localization is the task of identifying the direction of arrival (DOA) of a sound emitted by a source from a multichannel acquisition.
In this work, we leverage on SHD-LRA \cite{cobos2023acoustic}, a DOA estimation approach that exploits low-rank signal approximations in the spherical harmonic domain.
In particular, the algorithm makes use of the spherical harmonics representation in order to estimate the direction-dependent components that characterize the source position by means of low-rank decomposition of the expansion coefficients \cite{cobos2023acoustic}.

As in \cite{cobos2023acoustic}, results are evaluated in terms of two performance metrics: the probability of detection (PD) and the root mean squared error (RMSE) of the DOA.
In particular, the former is computed as the percentage of DOA estimates below an absolute DOA error of \SI{10}{\degree}.
Using the microphone signals captured by the Spatial Mics and considering the acoustically calibrated positions as ground truth, a probability of detection of \SI{79}{\percent} can be achieved.
This aligns with the results presented in \cite{cobos2023acoustic} when dealing with reverberant room conditions.
Also the RMSE values are consistent with those in \cite{cobos2023acoustic} computed for non-ideal conditions, yielding an azimuth RMSE of \SI{3.33}{\degree} and an elevation RMSE of \SI{6.10}{\degree}.

%% file: results/separation.tex
\definecolor{bblue}{HTML}{4F81BD}
\definecolor{rred}{HTML}{C0504D}
\definecolor{ggreen}{HTML}{9BBB59}

\begin{tikzpicture}
    \begin{axis}[
        width = 0.9\columnwidth,
        height = 4.7cm,
        major x tick style = transparent,
        ybar=2*\pgflinewidth,
        ymajorgrids = true,
        y unit = \decibel,
        symbolic x coords={Source $S_1$,Source $S_2$},
        xtick = data,
        scaled y ticks = false,
        enlarge x limits=0.55,
        ylabel near ticks,
        xlabel near ticks,
        ymin=0,
        legend cell align=left,
        legend style={
                at={(0.5,1.05)},
                anchor=south,
                column sep=1ex,
                legend columns=-1
        }
    ]
        \addplot[style={bblue,fill=bblue,mark=none}]
            coordinates {(Source $S_1$, 0.83) (Source $S_2$, 1.32)};
        \addplot[style={rred,fill=rred,mark=none}]
             coordinates {(Source $S_1$, 4.88) (Source $S_2$, 4.55)};
        \addplot[style={ggreen,fill=ggreen,mark=none}]
             coordinates {(Source $S_1$, 8.28) (Source $S_2$, 9.00)};
        \legend{SDR,SIR,SAR}
    \end{axis}
\end{tikzpicture}

%% file: sections/05_conclusion.tex
\section{Conclusion}
\label{sec:conclusion}

We presented HOMULA-RIR, a dataset of \acrshort{rirs} measured in a seminar room, acquired through the use of both ULAs and HOMs.
This diverse and versatile configuration guarantees suitability for a broad range of application scenarios in the context of telecommunications, teleconferencing, and spatial audio.
We provide precise measurements, together with acoustically calibrated microphone positions.
Various analyses, including the measurement of reverberation time and clarity, have been conducted to offer a comprehensive understanding of the acoustic characteristics within the environment.
To validate the usability of the dataset, we performed tests with two classic applications: blind source separation and source localization.
Results prove the effectiveness of the dataset in these contexts, showcasing its potential for the intended applications.

%% file: sections/06_ack.tex
\section{acknowledgements}
We would like to thank Voyage Audio LLC and Eventide Inc. for their support.